\documentclass[debug]{rmaa}

\usepackage{paralist}
\usepackage{amsmath}
\usepackage{natbib}

\usepackage{psfrag,color}

\usepackage[latin1]{inputenc}

\title{Ferrers bar response models: a grid calculation for Galactic models} 

\author{
  A. Silva-Castro,\altaffilmark{1} 
  and I. Puerari,\altaffilmark{1}}

\altaffiltext{1}{INAOE, M\'exico.}

\shortauthor{Silva-Castro \& Puerari}
\shorttitle{Ferrers bar response models}

\fulladdresses{
\item Silva-Castro, Alan \& Iv\^anio Puerari: Instituto Nacional de Astrof\'\i sica, \'Optica y Electr\'onica,
	Calle Luis Enrique Erro, No. 1, 72840, Santa Mar\'\i a Tonantzintla, Puebla, M\'exico
  (alan.silva@inaoep.mx, puerari@inaoep.mx).
    }

\listofauthors{A. Silva Castro \& I. Puerari}
\indexauthor{Silva-Castro, A.}
\indexauthor{Puerari, I.}

\abstract{This study numerically investigates the dynamics of barred spiral galaxies using 3D Ferrers bar response models. A total of 708 models were analyzed, incorporating variations in the axisymmetric potential (nucleus, bulge, disk, halo), bar length, mass, angular velocity, and disk stellar velocity dispersion. Model evaluation employed the Spearman correlation (to assess input-output relationships) and permutation feature importance in a Random Forest Regressor (to measure input variable impacts). Orbital configurations of test particles reveal the critical role of bar dynamics in shaping galaxies' morphological and kinematic properties. Key findings emphasize how bar potential influences major orbital families, affecting barred galaxies' long-term structure. These results provide deeper insights into galactic component interactions and a robust framework for understanding bar properties.}

\resumen{
Este estudio investiga numéricamente la dinámica de galaxias espirales barradas mediante modelos de respuesta de barras de Ferrers en 3D. Se analizaron 708 modelos, variando el potencial axisimétrico (núcleo, bulbo, disco, halo), la longitud, masa y velocidad angular de la barra, y la dispersión estelar del disco. La evaluación incluyó la correlación de Spearman y el ``permutation feature importance'' en un ``Random Forest Regressor''. Las configuraciones orbitales revelan el papel crítico de la barra en la formación de propiedades morfológicas y cinemáticas. Los resultados destacan cómo su potencial influye en las principales familias orbitales, afectando la estructura galáctica a largo plazo. Estos hallazgos permiten una comprensión más profunda de las interacciones galácticas y un marco sólido para estudiar las barras.
}

\addkeyword{Galaxies: evolution}
\addkeyword{Galaxies: fundamental parameters}
\addkeyword{Galaxies: kinematics and dynamics}
\addkeyword{Galaxies: statistics}
\addkeyword{Galaxies: structure}

\begin{document}
\maketitle

\section{Introduction}
\label{sec:intro}

Barred spiral galaxies are one of the most intriguing kind of objects in the universe. In these galaxies, a large bisymmetrical structure grows in the center of the disk component, modifying drastically the kinematics and the dynamics of the barions contained in the central part of the galaxy. Dark matter is also influenced by the formation of the bar structure \citep{2003MNRAS.345..406V}.\\

Since decades, barred galaxies are the subject of several observational, theoretical and numerical studies.  It is now well established that bars significantly influence their host galaxies in various ways. As the bar grows over time, it transfers angular momentum from the inner disk to the outer disk and the dark matter halo, as discussed by several authors (\citealp{1972MNRAS.157....1L}; \citealp{1981A&A....99..362S}; \citealp{2003MNRAS.341.1179A}; \citealp{2013MNRAS.429.1949A}). The growing bar will also direct gas to the center of the galaxy
along the narrow lanes that represent loci of shocks within the bar region (\citealp{1976Ap&SS..43..491S}; \citealp{1992MNRAS.259..345A}; \citealp{2004A&A...416..515D}; \citealp{2010ApJ...719.1470V}; \citealp{2017MNRAS.465.3729S}; \citealp{2019A&A...621L...4G}) triggering the nuclear activity. The influence of the bar on gas dynamics has been the subject of extensive research over the years, with a wide range of studies dedicated to this topic (E.g. \citealp{1981ApJ...246..740V}; \citealp{1985MNRAS.212..677S}; \citealp{1995ApJ...449..508P}; \citealp{2002MNRAS.329..502M};
\citealp{2012ApJ...751..124K}; \citealp{2022Univ....8..290P}; \citealp{2024MNRAS.528.5742S}) and references therein.\\

There are also studies proving the kinematics of bars to be significant. In numerical simulations, the bar pattern speed ($\Omega_B$), or angular velocity of the bar, is closely linked to the evolution of both the bar and its host galaxy. As the bar grows and transfers angular momentum to its surroundings, it generally slows down, resulting in a decrease in the pattern speed (\citealp{2000ApJ...543..704D}; \citealp{2003MNRAS.341.1179A}; \citealp{2006ApJ...637..214M}; \citealp{2015PASJ...67...63O}; \citealp{2018ApJ...860..152W}). The bar evolution is also influenced by its surrounding environment. Recent studies have demonstrated that the angular momentum of the dark matter halo plays a crucial role in shaping the evolution of both the bar and the disk, impacting the bar pattern speed, instability timescales, and other dynamics (\citealp{2013MNRAS.434.1287S}; \citealp{2016MNRAS.463.1952P}; \citealp{2018MNRAS.476.1331C}; \citealp{2021ApJ...915...23C}). \\

Observational studies have shown that barred galaxies exhibit increased star formation in their central regions (\citealp{1977Natur.266..607M}; \citealp{1986MNRAS.221P..41H}; 
\citealp{1991RMxAA..22..197G}; \citealp{1992ApJ...395L..79K}; \citealp{2001AJ....122.1350A}; \citealp{2008A&A...482..133H}; \citealp{2011ApJ...743L..13C}; \citealp{2011MNRAS.416.2182E}; \citealp{2020MNRAS.499.1406L}; \citealp{2024ApJ...973..129G}) as well as in the bar-end regions (\citealp{1998A&A...337..671R}; \citealp{2007A&A...474...43V}; \citealp{2020A&A...644A..38D}; \citealp{2020MNRAS.495.4158F}; \citealp{2020MNRAS.495.3840M}; \citealp{2024ApJ...973..129G}). Conversely, star formation is suppressed along the arms of the bar (\citealp{1998A&A...337..671R}; \citealp{2004A&A...413...73Z}; \citealp{2011MNRAS.411.1409W}; \citealp{2016A&A...589A..66H}; \citealp{2024ApJ...973..129G}). These observational and numerical studies highlight the significant role that bars play in the evolution of their host galaxies.\\

In this paper, we present a new numerical investigation by studying a large number of three-dimensional  response models. The axisymmetric part of the models is generated to fit the Galactic circular rotation curve proposed by \citet{2020Galax...8...37S}. We tested a number of parameters as disk particles velocity dispersion, and mass, size and angular pattern speed of the imposed bar. In section \ref{sec:Simul} we present our models, parameters, initial conditions and orbits integration, and in Section \ref{sec:property_bar} we discuss the resulting bar structure observed in the particles distribution. The resulting values are discussed in Section \ref{sec:feat_res} and finally in Section \ref{sec:conclu} we present a general discussion and our conclusions.

\section{Simulations}
\label{sec:Simul}

\subsection{Galactic models}
\label{sec:GM}

\begin{figure}[!t]
  \includegraphics[width=\columnwidth]{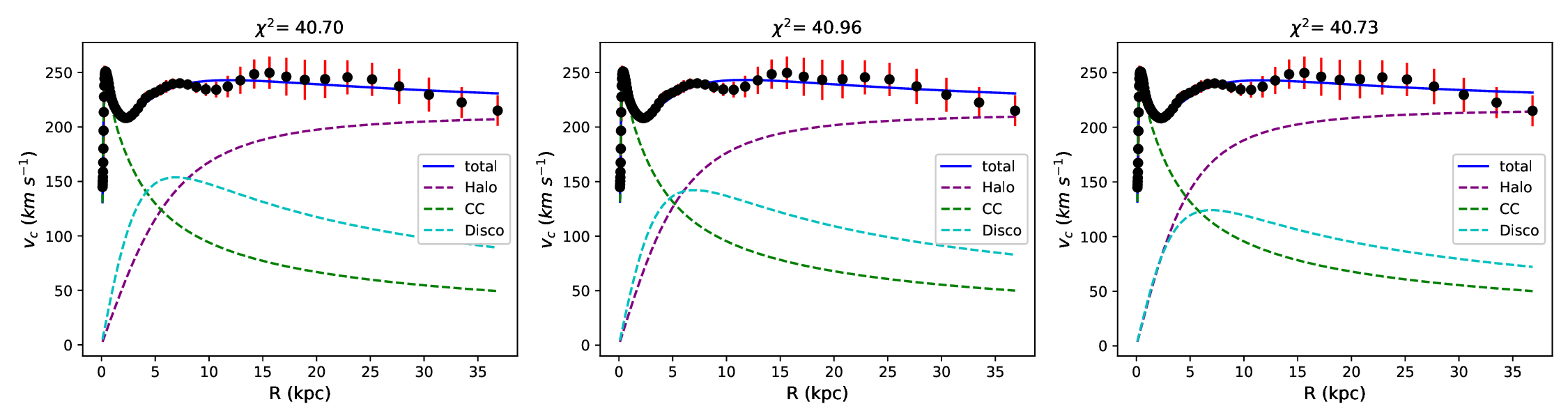}
  \caption{The circular velocity curves corresponding to the axisymmetric potential for three different sets of parameter values. CC means Central Component (bulge$+$core). These curves are compared with the circular velocity of the Milky Way \citep{2020Galax...8...37S} within the first 40 kpc, along with their associated $\chi^2$.}
  \label{im:vel_circ}
\end{figure}

The selected gravitational potentials are composed by a sum of the axisymmetric and bar components. The axisymmetric component itself is a superposition of several elements: a core and a bulge, both modeled by a Plummer potential \citep{1911MNRAS..71..460P}, a disk represented by a Miyamoto-Nagai model \citep{1975PASJ...27..533M}, and a halo described by a logarithmic potential \citep{1980ApJ...238..103R}. The full axisymmetric potential is then

\begin{equation}
    \Phi_{ax}=  -\frac{GM_{c1}}{r^2+r_{c1}^2}-\frac{GM_{c2}}{r^2+r_{c2}^2}-\frac{GM_D}{\sqrt{R^2 + (a + \sqrt{z^2 + b^2})^2}}+\frac{v_H^2}{2} \ln(r^2 +r_H^2)
\end{equation}

\noindent where $r^2=x^2+y^2+z^2$, while $R^2=x^2+y^2$.
$M_{c1}$ and $M_{c2}$ are the masses of the bulge and the core, respectively, and $r_{c1}$ and $r_{c2}$ are their radial scale lengths. $M_D$ is the disk mass and $a$ and $b$ its structural parameters. For the halo, $v_H$ is the asymptotic velocity and $r_H$ its radial scale length.
By comparing the circular velocity profiles obtained from different parameter sets with the observed velocities in the Milky Way within the first 40 kpc, as reported by \citet{2020Galax...8...37S}, and by employing the gradient descent method, we identified three parameter sets with nearly identical $\chi^2$ values. This indicates that the potentials corresponding to these parameters are degenerate. However, these potentials exhibit distinct characteristics from one another. Figure \ref{im:vel_circ} displays the circular velocities derived from the three models, which we will henceforth refer to as Model 1 (left panel), Model 2 (middle panel), and Model 3 (right panel), based on the influence of their respective disks. The parameter values corresponding to each model are provided in Table \ref{tab:par_axy}.

\begin{table}[!t]\centering
  \setlength{\tabcolsep}{0.8\tabcolsep}
  \caption{Parameters for the axisymmetric models} \label{tab:par_axy}
 \resizebox{\columnwidth}{!}{\begin{tabular}{lccccccccc}
    \toprule
    \textbf{Model} & $\mathbf{r_{c1} \, \, (kpc)}$ & $\mathbf{M_{c1} \, \, (M_\odot)}$ & $\mathbf{r_{c2} \, \, (kpc)}$ & $\mathbf{M_{c2} \, \, (M_\odot)}$ & $\mathbf{a \, \, (kpc)}$ & $\mathbf{b \, \, (kpc)}$ & $\mathbf{M_{D} \, \, (M_\odot)}$ & $\mathbf{r_{H} \, \, (kpc)}$ & $\mathbf{v_{H} \, \, (km/s)}$ \\ 
\textbf{1} & $1.46$ & $1.07\times 10^{10}$ & $0.278$ & $1.01\times10^{10}$ & $4.66$ & $0.233$ & $6.99\times 10^{10}$ & $7.68$ & $212$ \\ 
\textbf{2} & $1.46$ & $1.15\times 10^{10}$ & $0.275$ & $1.00\times10^{10}$ & $4.72$ & $0.236$ & $6.06\times 10^{10}$ & $6.78$ & $213$ \\ 
\textbf{3} & $1.45$ & $1.16\times 10^{10}$ & $0.274$ & $9.96\times10^{9}$ & $4.69$ & $0.235$ & $4.60\times 10^{10}$ & $5.73$ & $217$ \\ 
    
    \bottomrule
  \end{tabular}}
\end{table}

\subsection{Bar Models}
\label{sec:BM}

For the bar component, we employed the ellipsoidal Ferrers potential. The mass density associated with this potential is defined as:

\begin{equation}
    \rho_B(m) =
    \begin{cases}
    \rho_{B_c}(1-m^2)^n & \text{for } m \leq 1 \\
    0 & \text{for } m > 1 \\
    \end{cases}
    \label{eq:ferrer}
\end{equation}

\noindent where $\rho_{B_c}=\frac{105}{32\pi}\frac{M_B}{a_B b_B c_B}$ is the central density and $m=(x/a_B)^2+(y/b_B)^2+(z/c_B)^2$, while $a_B$, $b_B$ and $c_B$ are the semi-axes of the ellipsoid. The index $n$ is chosen to be $n=2$. The forces generated by this potential are described in \citet{1984A&A...134..373P}.\\

To simplify the models, we fixed the ratios between the semi-axes, setting $b_B=a_B/3$ and $c_B=a_B/6$. However, the major semi-axis $a_B$ was varied, considering three distinct values: 6, 4.5, and 3 kpc. We introduced the bar component as a smooth time-dependent function by gradually transferring mass from the disk to the bar in the following way:

\begin{equation}
     M_B(t) =
    \begin{cases}
        \dfrac{M_{B_f}}{2}\bigg(1-\cos\bigg(\dfrac{\pi}{T_{max}}t \bigg) \bigg) & \text{for } 0 \leq t \leq T_{\text{max}} \\
        M_{B_f} & \text{for } t > T_{\text{max}} \\
    \end{cases}
    \label{eq:mass_time}
\end{equation}

\noindent Here, $ M_{B_f}$ represents the final mass of the bar after its growth is completed at $T_{\text{max}}$ which is set to 1 Gyr out of a total simulation time of 11.25 Gyr. Then, our models have a transient phase of 1 Gyr and are time independent after that. The evolution of the disk mass is then expressed as  $M_D(t) = M_{D_i} - M_B(t)$. Rather than directly referring to $M_{B_f}$, we will use  $\mu_B = \frac{M_{B_f}}{M_{D_i}}$, which denotes the final fraction of mass transferred from the disk to the bar.\\

The parameter $\mathcal{R} = \frac{R_{CR}}{a_B}$ ($R_{CR}$  is the corotation radius, i.e., the radius at which the stars have the same angular speed as the pattern speed of the bar) is used to characterize the bar rotation rate. Bars are kinematically classified as `slow' if $\mathcal{R} > 1.4$, `fast' if  $1.0 < \mathcal{R} < 1.4$ or `ultrafast' if $\mathcal{R} < 1.0$ (see  \citet{2000ApJ...543..704D}; \citet{2008MNRAS.388.1803R}; 
\citet{2015A&A...576A.102A}; \citet{2022ApJ...926...58L}).
Concerning the $\mathcal{R}$ found for our barred models, we must note that each of the models have different axisymmetric backgrounds (Model 1, 2 or 3), different bar semi major axis $a_b$, different $\mu_B = \frac{M_{B_f}}{M_{D_i}}$ and different $\Omega_B$. However, after the bar growth phase, with all parameters set, $\mathcal{R}$ depends solely on ${R_{CR}}$ (determined by $\Omega_B$) and $a_B$. \\

\begin{figure}[!t]
  \includegraphics[width=\columnwidth]{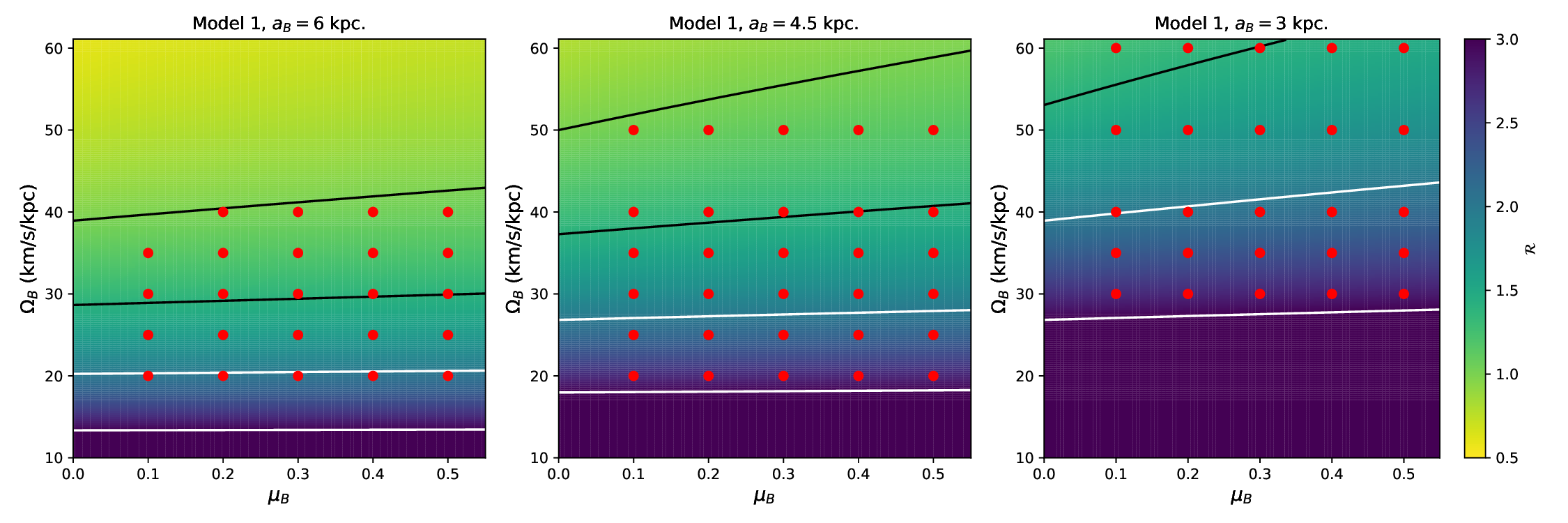}
  \caption{Parameter space for $\mathcal{R}$ with respect of the parameters $\mu_B$ and $\Omega_B$ for model 1 and three values of bar length. The red dots are the models we studied in this article. Here we show the values only for Model 1, because there is not a significant change of $\mathcal{R}$ when considering the three axisymmetric models (Model 1, Model 2 and Model 3), once they are time independent (see text). From bottom to top, the straight lines in each plot denote equal $\mathcal{R}$: 3 and 2 (white solid lines) and 1.4 and 1 (black solid lines).
  }
  \label{im:vel_bar}
\end{figure}

We are now able to analyze the bar rotation rate in a two-dimensional parameter space for $\mathcal{R}$ corresponding to each axisymmetric model, alongside the three selected values for the bar length, as illustrated in Figure \ref{im:vel_bar} for Model 1. This methodology allows for the identification of models featuring either fast or slow bars, with varying values of $\mu_B$ and $\Omega_B$. Therefore, we focused on models with $\mu_B = 0.1, 0.2, 0.3, 0.4, 0.5$ and $\Omega_B = 20, 25, 30, 35, 40, 50, 60 \ \text{km/s}.$ Furthermore, only models with $1 \leq \mathcal{R} \leq 3$ were analyzed (indicated by red dots in Figure \ref{im:vel_bar}), corresponding to bars classified as fast and slow by \citet{2015A&A...576A.102A}.

\subsection{Initial conditions and orbits integration}
Given our focus on stars within the disk, we selected the initial positions of the test particles following the axisymmetric distribution of the Kuzmin model which surface density is \citep{2008gady.book.....B}:

\begin{equation}
    \Sigma_K(R)=\frac{aM_D}{2\pi(R^2+a^2)^{3/2}}
    \label{eq:dens_sup}
\end{equation}

We have chosen to use KT models to create the initial particle distribution since our Miyamoto-Nagai models have scale length ratio ($a/b$) of 20, i.e. the disks are very thin. Furthermore, all the particles are initially in the galactic plane ($z(t=0) = 0$).  Considering the relationship between surface density and the number of stars, we observe that at a given radius, the number of stars can be expressed as $N_K(R)=2\pi \Sigma_K(R)$. By employing the Monte Carlo method, we can achieve a well-distributed arrangement of stars within the galactic plane. The maximum radius selected  was given by $R_{max}=1.5 R_{CR}$. Although all particles are initially positioned within the galactic plane, the three-dimensional aspect of the simulations is established through the introduction of an initial velocity component along the $z$-axis.\\

For the velocity initial conditions, we decided to make $\Bar{v}_\theta=v_c=\sqrt{R{d\Phi_{ax} \over dR}}$, while $\Bar{v}_R=\Bar{v}_z=0$, since all the stars in an axisymmetric galaxy (as in our models for $t=0$) have near-circular orbits. On the other hand, the velocity dispersion for each coordinate decreases exponentially with radius as noted by \citet{1989AJ.....97..139L}:

\begin{equation} \label{eq_disps}
    \begin{aligned}
        \sigma_{R}(R)=&\sigma_R(0)\exp{\bigg(-\frac{R}{2h_R}\bigg)}\\
        \sigma_{T}(R)=&\sigma_T(0)\exp{\bigg(-\frac{R}{2h_T}\bigg)}\\
        \sigma_{z}(R)=&\sigma_z(0)\exp{\bigg(-\frac{R}{2h_z}\bigg)}\\
    \end{aligned}
\end{equation}

\noindent where $\sigma_{R}$, $\sigma_{T}$ and $\sigma_{z}$ are the radial, tangential and vertical velocity dispersions, $\sigma_R(0)$, $\sigma_T(0)$ $\sigma_z(0)$ are their central values,  and $h_R$, $h_T$ and $h_z$ are the scale lengths. In their work, \citet{1989AJ.....97..139L} computed the specific values of $h_R$ and $h_T$ for the Milky Way, which were found to be 4.37 kpc and 3.36 kpc, respectively. In the same work, Lewis and Freeman also assumed that the ratio of radial velocity dispersion $\sigma_R$ to vertical velocity dispersion $\sigma_z$ remains constant. Consequently, we have chosen $h_z=h_R$.\\

 For our research, we have chosen to equate the three central velocity dispersions ($\sigma_R(0)=\sigma_T(0)=\sigma_z(0)=\sigma_D$). Additionally, we explore three distinct values for this new parameter: 100, 80 and 50 km/s.\\

Hence, in total, we have 5 different galactic parameters: three different axisymmetric models, five distinct bar/disk mass ratios, seven angular velocities, three set of bar lengths and three disk central velocity dispersion. This resulted in 945 different models. However, as mentioned before, only models with $1 \leq \mathcal{R} \leq 3$ were studied and then, our final set of simulations  discussed here included 708 simulations. Each simulation comprising a total of 30,000 test particles, yielding a total of 21,240,000 calculated orbits. A concise summary of the models parameter space is presented in Table \ref{tab:par_space}.\\

\begin{table}[!t]\centering
  \setlength{\tabnotewidth}{1\columnwidth}
  \tablecols{2}
  \caption{Models parameter space.} \label{tab:par_space}
 \resizebox{\columnwidth}{!}{\begin{tabular}{lc}
    \toprule
    \textbf{Parameter} & \textbf{Value} \\ \hline
Model & 1, 2, 3 \\
$\sigma_D$ (km/s)  & 50, 80, 100 \\ 
$a_B$ (kpc)  \tabnotemark{a} & 3, 4.5, 6 \\
$\mu_B$  & 0.1, 0.2, 0.3, 0.4, 0.5 \\
$\Omega_B$ (km/s/kpc) & 20, 25, 30, 35, 40, 50, 60 \\
    
\bottomrule

\tabnotetext{a}{The axis ratios are $b_B=a_B/3$ and $c_B=a_B/6$.}
\end{tabular}}
\end{table}

The integration was carried out using a fourth-order Runge-Kutta integrator, employing Fortran subroutines. To assess stability, we monitored the Jacobi energy of test particles following bar growth. Typically, $\Delta E_J$ is better than $10^{-10}$ for $t>T_{max}$, since only then the models become time independent. The total simulation time was 11.25 Gyr, during which 1225 snapshots were captured at equidistant intervals. The mass bar growth ceases at $T_{max}=1$ Gyr to ensure 1024 snapshots after the bar mass evolution. This number of points ($2^{10}$) was chosen to simplify the Fast Fourier Transform analysis.

\section{Estimating properties of the bar}
\label{sec:property_bar}

\subsection{Detecting the periodic orbits $x_1$ and $x_2$}
\label{sec:fam}

As highlighted by several authors (e.g., \citet{1970AJ.....75...96C},
\citet{1983A&A...127..349A},
\citet{1993RPPh...56..173S},
\citet{2002MNRAS.333..847S} and \citet{2019MNRAS.490.2740P}), periodic orbits play a crucial role in shaping the structure of barred galaxies. These periodic orbits are categorized into four primary families, namely the $x_1$, $x_2$, $x_3$ and $x_4$ family orbits, following the classification by \citet{1980A&A....92...33C}, with the most significant being the  $x_1$ and $x_2$ families. Identifying these periodic orbits is essential for understanding the dynamics of barred galaxies.\\

To detect these orbits in our simulations, we applied a Fourier transform on the particle coordinates projected onto the equatorial plane: $x(t)$, $y(t)$, and $R'(t) = R(t) - \bar{R}$, where $\bar{R}$ is the mean radius of the particle orbit. This allowed us to extract the dominant frequencies, i.e., those with the highest amplitudes, in each coordinate. These frequencies are  denoted as $\omega_x$, $\omega_y$, and $\omega_R$, respectively. Additionally, the corresponding amplitudes were determined and labeled as $A_x$, $A_y$, and $A_R$.\\

It is crucial to emphasize that this analysis was conducted after the bar has reached its final mass, i.e., for $t > T_{max}$, as in our approach the stellar orbits stabilize and maintain a consistent pattern after the bar growth. In addition, prior to performing the Fourier transform, we applied a Hanning window function to the orbits positions. The purpose of this window function was to mitigate signal `leakage' in the Fourier spectra.\\

We can now identify sticky orbits around the $x_1$ and $x_2$ families of periodic orbits (\citealp{2008IJBC...18.2929C}; \citealp{2013IJBC...2330005K}), but for simplicity, we will refer to them as members of the $x_1$ or $x_2$ families. An orbit is classified as part of the \( x_1 \) family if it satisfies the condition $1.9 \leq \omega_R / \omega_x \leq 2.1$ and $A_x / A_y \geq 2$. On the other hand, orbits that meet the criteria $1.9 \leq \omega_R / \omega_x \leq 2.1$ and $A_x / A_y \leq 0.5$ are associated with either the $x_2$ or $x_3$ family. However, it is important to note that the $x_3$ family is significantly less stable than the $x_2$ family \citep{2002MNRAS.333..861S}, and as such, the presence of sticky-chaotic orbits around $x_3$ is expected to be minimal, making them insignificant for classification purposes.\\

Having identified the orbits belonging to the $x_1$ and $x_2$ families, we can now quantify the number of orbits in each family, denoted as $N_{x_1}$ and $N_{x_2}$, respectively. Alternatively, we can calculate the proportion of  $x_1$ and $x_2$ orbits relative to the total number of elliptical orbits, i.e., orbits that satisfy $1.9 \leq \omega_R / \omega_x \leq 2.1$. These proportions are denoted as $P_{x_1}$ and $P_{x_2}$, respectively.

\subsection{Bar strength}
\label{sec:Bar_s}

We also performed a Fourier analysis on the stars positions to calculate the bar strength. For this, we computed the m=2 mode Fourier coefficients ($a_2$ and $b_2$) based on the particles position located within an annulus of width $\Delta R$ at a radius $R$. Hence, as highlighted by \citet{2024ApJ...965...77C}, the amplitude of the bar is:

\begin{equation}
    \tilde{A}^2_2(R)=a_2^2+b_2^2
    \label{eq:bar_amp}
\end{equation}

\noindent where the bar strength corresponds to the maximum value $\tilde{A}_2$ within $R_{CR}$, expressed as:

\begin{equation}
    A_2\equiv \mathop{\text{max}}_{R<R_{cor}}\tilde{A}_2(R)
    \label{eq:A2}
\end{equation}

\begin{figure*}[!t]
  \includegraphics[width=0.90\columnwidth]{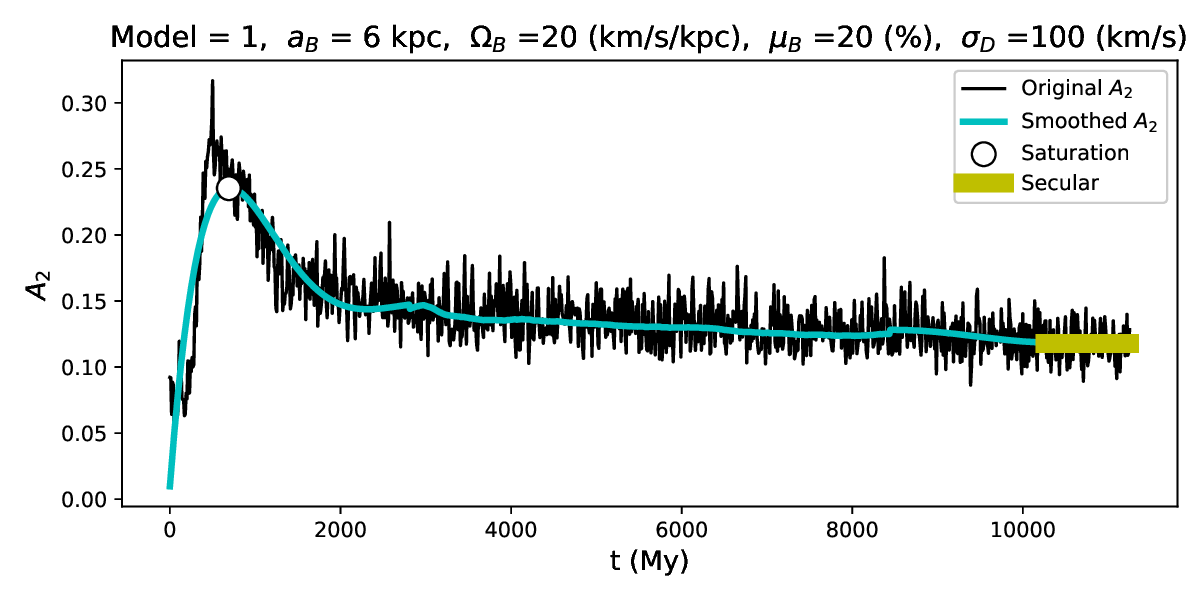}%
  \hfill
  \includegraphics[width=0.90\columnwidth]{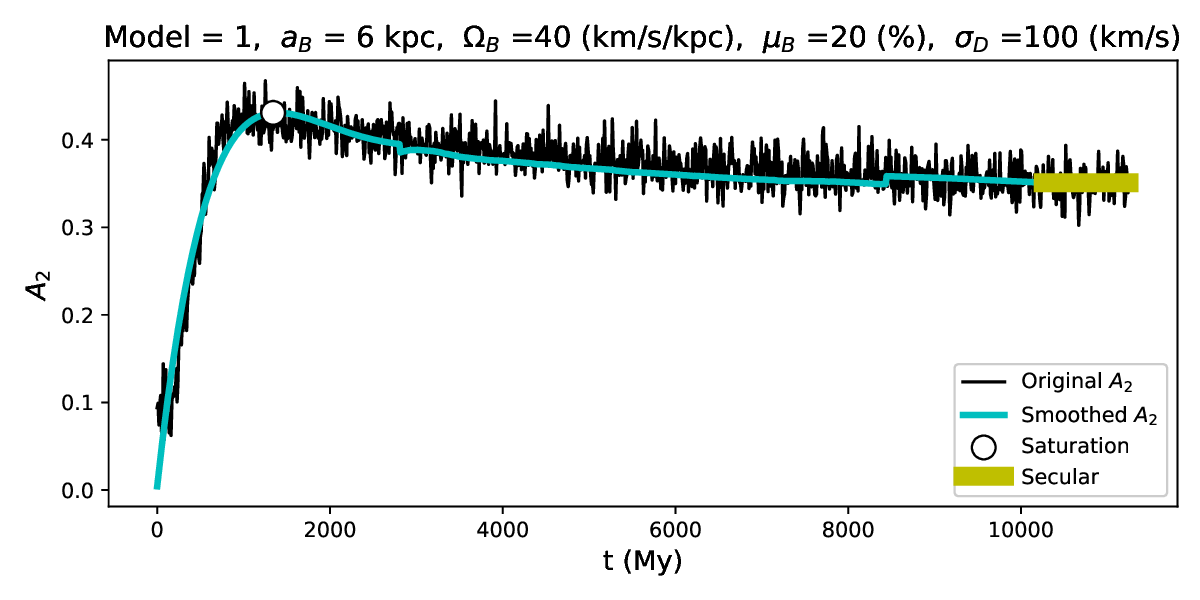}
  \caption{Evolution of the bar strength $A_2(t)$ over time (black line) for two different models. The cyan line corresponds to the smoothed versions. Additionally, the white/black dot indicates the point of saturation in $A_2(t)$, while the horizontal yellow line represents the secular value associated. Notice the different behavior of $A_2(t)$; in one simulation, $A_2(t)$ decreases after the time of saturation, while in the other model, $A_2(t)$ remains more or less constant.}
  \label{im:A2_OG}
\end{figure*}

We applied this method to each snapshot, making $A_2$ a time-dependent parameter $A_2(t)$. This time evolution of $A_2(t)$ was then used to analyze the properties and dynamics of the orbits in our galaxy models. In Figure \ref{im:A2_OG}, we observe the temporal evolution of $A_2(t)$ for two given simulations. In the same figure, a smoothed version of this quantity is shown (the smoothing process employed a Savitzky-Golay filter with a sixth-order polynomial).\\

Even with the smoothed curves of $A_2(t)$, analyzing each curve individually is impractical since we have a very large number of galaxy models; therefore, we seek for characteristic values that help us to evaluate the model.\\

One such value is the saturation point in $A_2(t)$. As observed in Figure \ref{im:A2_OG}, the $A_2(t)$ curves initially experience rapid growth, followed by a decline. Notably, several curves exhibit this decline after a specific time. We designate the values at this inflection point as $A_{sat}$ and $t_{sat}$.\\

Another characteristic value becomes evident toward the end of the simulations for $A_2(t)$. At this point, the curves exhibit minimal changes over time. This behavior is expected since the structure of rotating Ferrers bars is primarily supported by the stable portion of the $x_1$ family. In a response model, any changes in $A_2$ after a couple of bar revolutions beyond $T_{\text{max}}$ can therefore be attributed to the influence of chaotic or escaping orbits. To quantify this stability in the $A_2$ values, we calculated the average values over the last 1 Gyr. Specifically, we denote these values as  $<A_{sec}>$.\\

An additional observation is the difference between $A_{sat}$ and $<A_{sec}>$. Consequently, we designate this difference as another characteristic value, denoted by $\Delta A = A_{sat} - <A_{sec}>$.

\section{Feature parameters vs result values}
\label{sec:feat_res}

Having obtained several parameters that characterize the orbital behavior in each galaxy model, namely: $N_{x_1}$, $N_{x_2}$, $P_{x_1}$, $P_{x_2}$, $A_{sat}$, $t_{sat}$, $<A_{sec}>$, and $\Delta A$, collectively referred to as the `result values', we can now proceed to compare these values with the parameters that describe the galaxy, hereafter called the `feature parameters'. These feature parameters include: Model, $a_B$, $\Omega_B$, $\mu_B$, and $\sigma_D$. Additionally, we will incorporate the parameter $\mathcal{R}$ previously introduced, along with the quadrupole moment of the bar as calculated by \citet{2021MNRAS.502.4708G} for $n=2$ (Eq. \ref{Eq:quadru}) and the bar angular momentum for $n=2$ (Eq. \ref{Eq:mom_ang}).

\begin{equation}
    Q_B=\frac{M_B}{9}\bigg(a_B^2-b_B^2 \bigg)
\label{Eq:quadru}
\end{equation}

\begin{equation}
    L_B=\frac{M_B \Omega_B}{9}(a_B^2+b_B^2)
    \label{Eq:mom_ang}
\end{equation}

\begin{figure}[!t]
  \includegraphics[width=\columnwidth]{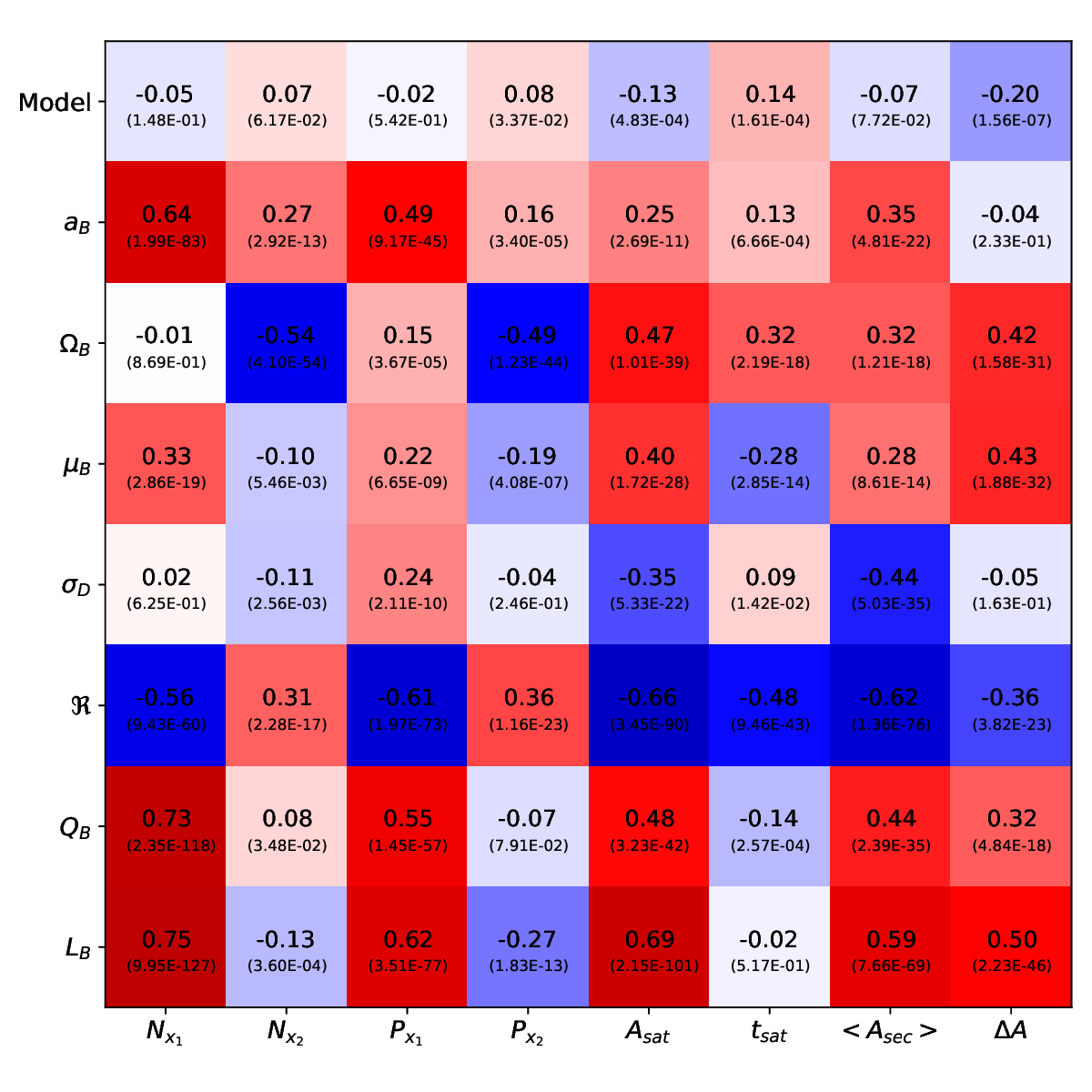}
  \caption{Spearman correlation coefficients between the result values and the feature parameters for all of our models (708 models). The associated p-values are provided in parentheses. Bright red colors represent high correlations, while bright blue ones represent high anti-correlations. The lighter the color, the weaker the correlation of a given feature parameter with the result value.}
  \label{im:spearman}
\end{figure}

We employed two methods to compare the result values and the feature parameters. The first method involves calculating Spearman correlation coefficients \citep{spearman}, which measure the strength and direction of a monotonic relationship between two ranked variables. This approach allows us to observe how the feature parameters affect the result values. The coefficients, along with their corresponding p-values (which represent the probability of obtaining test results at least as extreme as the result actually observed \citep{spearman}) are presented in Figure \ref{im:spearman}.\\

The second method, derived from machine learning, uses permutation feature importance within a Random Forest Regressor (RFR). An RFR \citep{2001MachL..45....5B} is an ensemble learning method that combines multiple decision trees to improve predictive performance and reduce overfitting.\\

Initially applied in this field by \citet{2021MNRAS.502.4708G}, this approach evaluates the contribution of each feature to a given result value by randomly shuffling the values of a specific feature and measuring the resulting change in the so called $R^2$ score. The $R^2$ score quantifies how well the model explains the variance in the result value, ranging from 1 (perfect fit) to $-\infty$ (arbitrarily poor fit) \citep{2011JMLR...12.2825P}. The difference between the $R^2$ score for the original data ($R^2_{\text{baseline}}$) and the permuted data ($R^2_{\text{permuted}}$) reveals the feature's contribution to the model's performance (\citealp{2021MNRAS.502.4708G}; \citealp{2001MachL..45....5B}). This method helps identify the most influential features, facilitating effective feature selection.\\

We trained the RFRs using 80\% of the data, reserving the remaining 20\% for testing, to find the optimal number of features for each result value. We set the number of trees in the model to 1,000 and determined the optimal model by selecting the one with the highest $R^2$. After identifying the best parameters for each result value, we retrained the models using the entire dataset, following the method proposed by \citet{2021MNRAS.502.4708G}.\\

\begin{figure}[!t]
	\includegraphics[width=0.50\columnwidth]{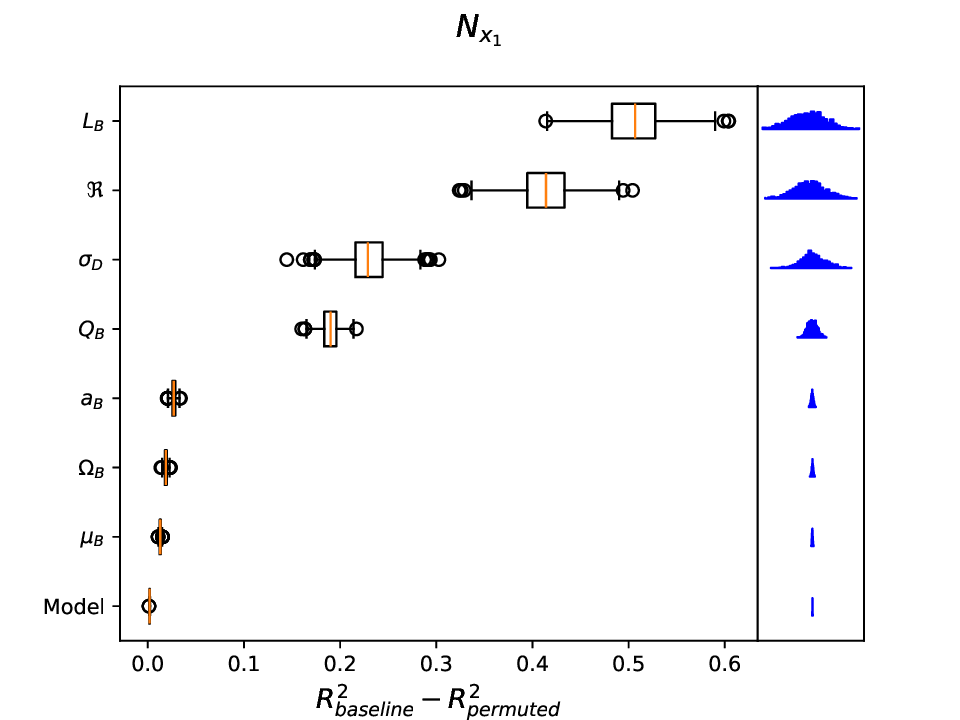}\hspace{0em}%
	\includegraphics[width=0.50\columnwidth]{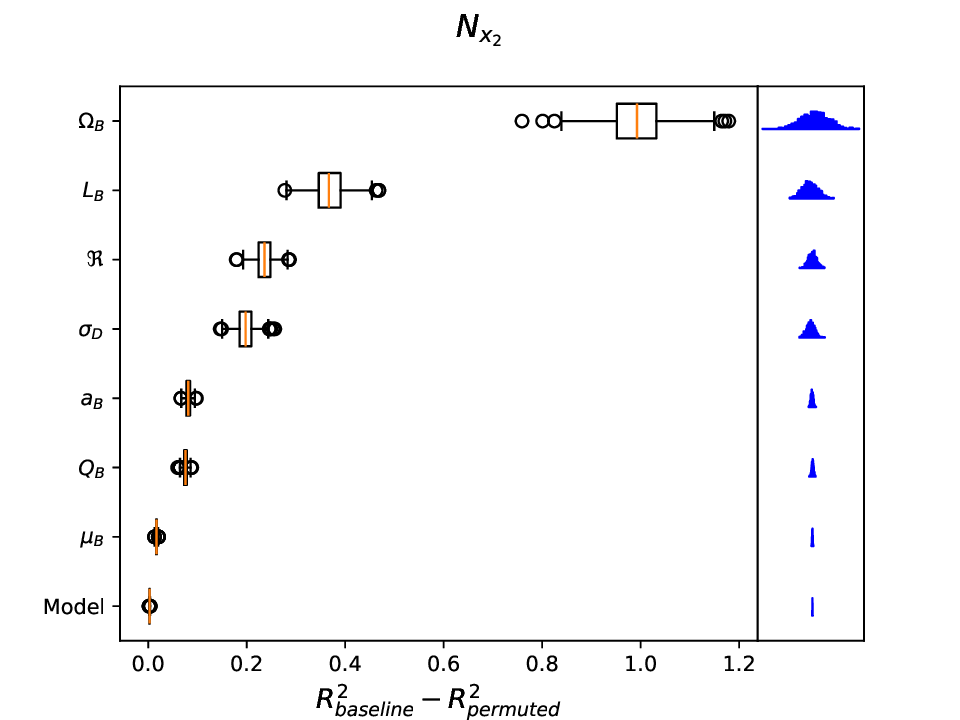}\hspace{0em}%
	\includegraphics[width=0.50\columnwidth]{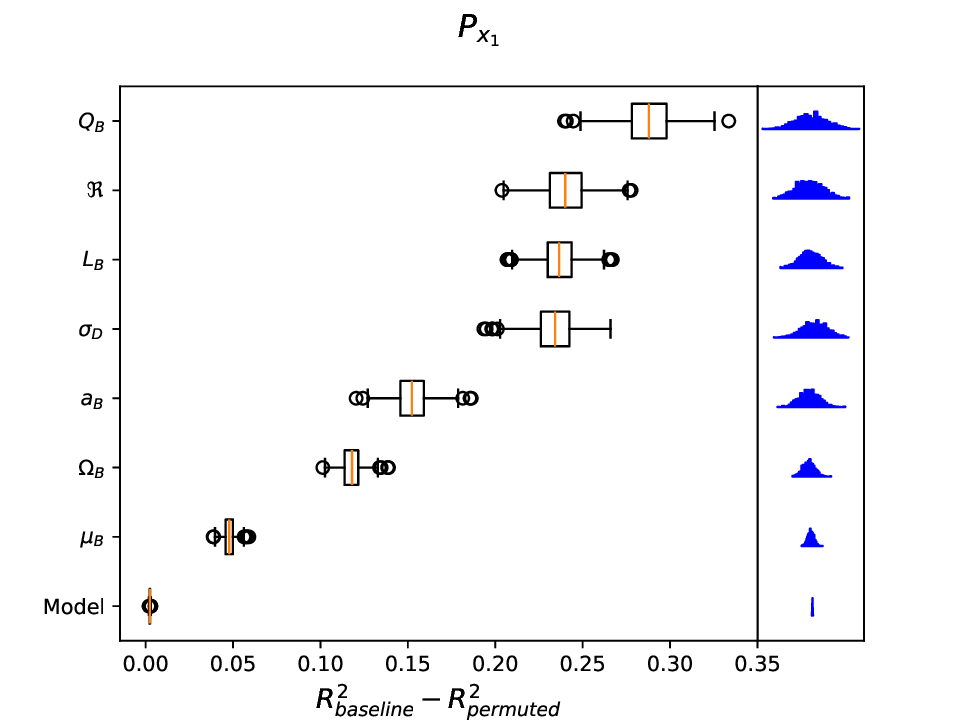}\hspace{0em}%
	\includegraphics[width=0.50\columnwidth]{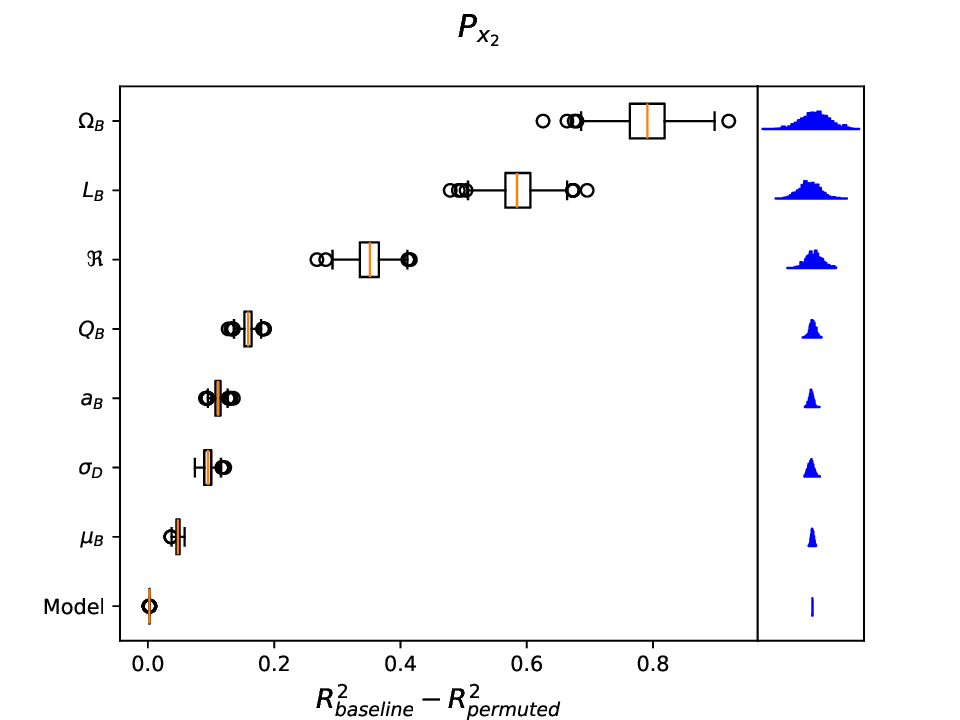}
	\caption{The permutation feature importance calculated using RFRs, trained to predict each specific result value related to the $x_1$ and $x_2$ families, is depicted. Each feature was permuted 1000 times, generating a distribution of $R^2_{baseline} - R^2_{permuted}$ scores. The orange line represents the median value of the distribution. The box limits indicate the 25th and 75th percentiles, while the whiskers extend to the minimum and maximum values. Outliers in the distribution are shown as open circles. The distribution itself is illustrated on the right side for each feature parameter.}
  \label{im:importances1}
\end{figure}

\begin{figure}[!t]
	\includegraphics[width=0.50\columnwidth]{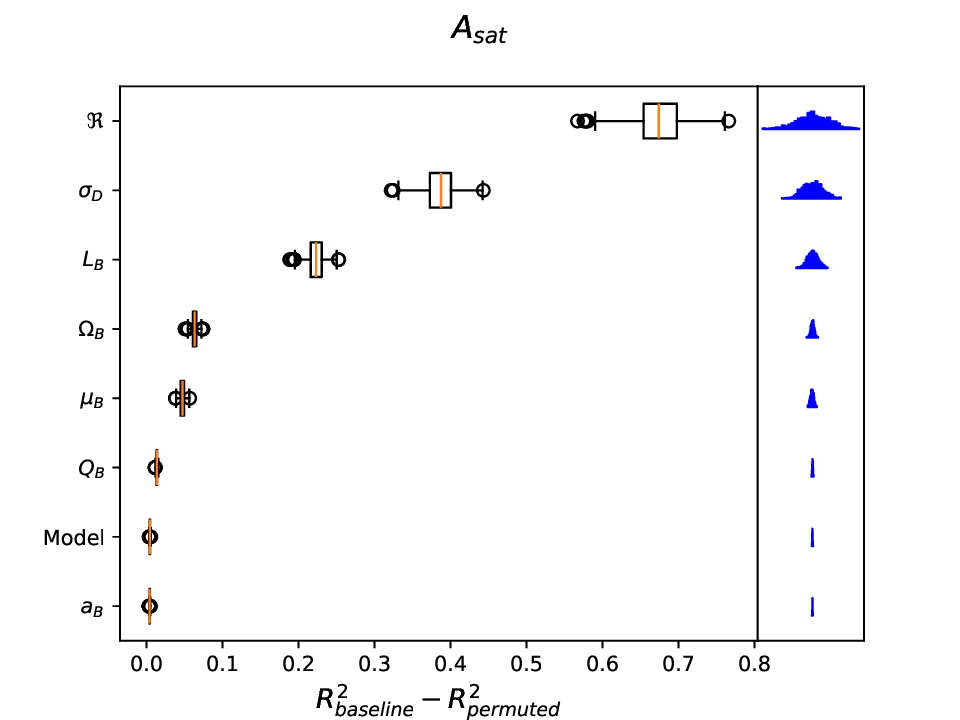}\hspace{0em}%
	\includegraphics[width=0.50\columnwidth]{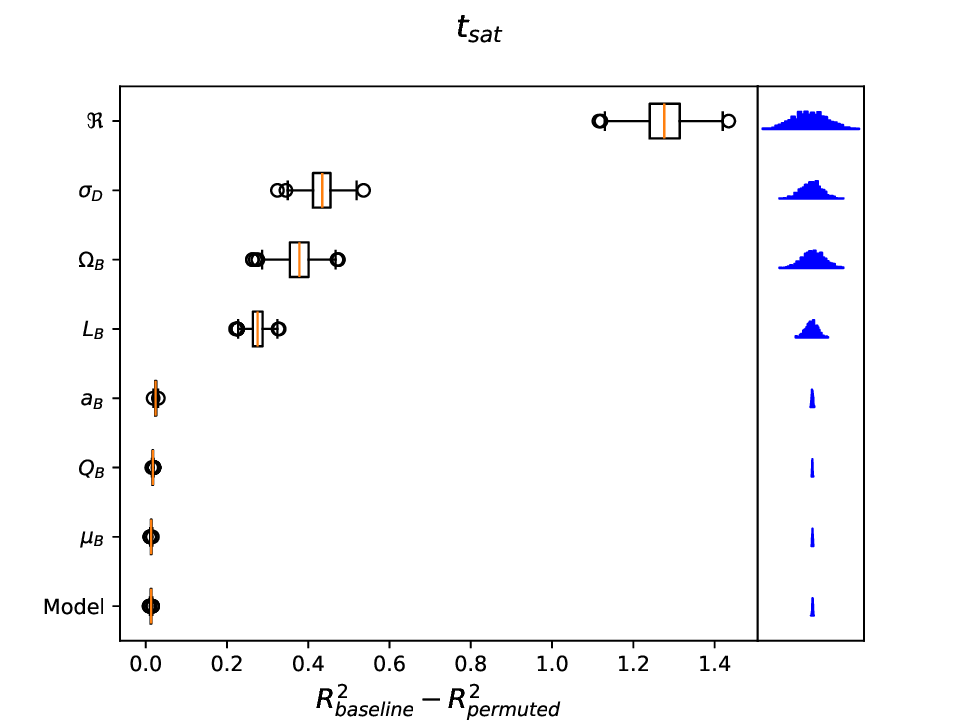}\hspace{0em}%
	\includegraphics[width=0.50\columnwidth]{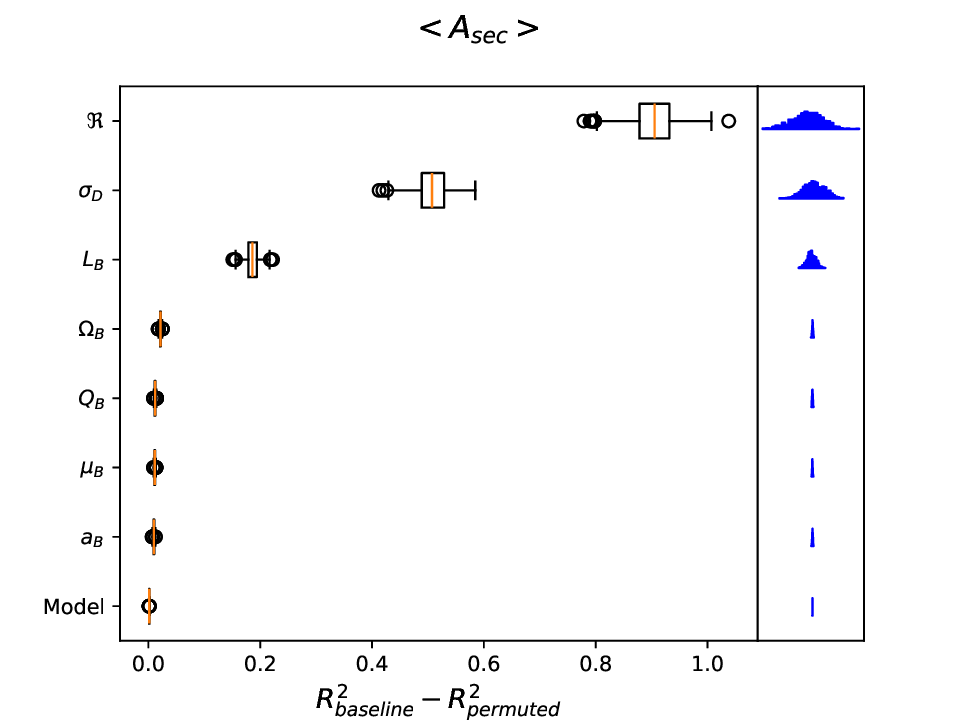}\hspace{0em}%
	\includegraphics[width=0.50\columnwidth]{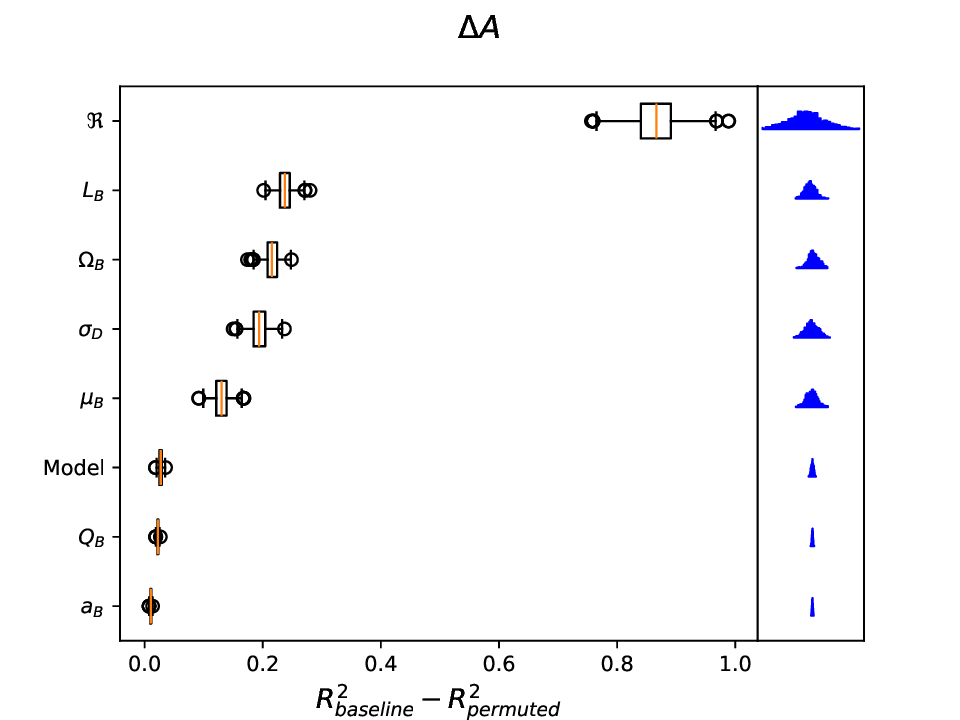}
	\caption{As Figure \ref{im:importances1} but for the result values related to the bar strength.}
	 \label{im:importances2}
\end{figure}

Finally, we estimated the permutation feature importance after 1,000 permutations for all result values, with the outcomes illustrated in Figures \ref{im:importances1} and \ref{im:importances2}, along with the feature parameter distributions. It is crucial to emphasize that the importance values, represented by $R^2_{baseline}-R^2_{permuted}$, are not absolute. Consequently, these values should not be compared across different result values, but rather among feature parameters within the same result value.\\

As previously noted, we are now able to assess the relative importance of each feature parameter for a given result value through the use of feature importance, as illustrated in Figures \ref{im:importances1} and \ref{im:importances2}. Additionally, the Spearman correlation coefficients presented in Figure \ref{im:spearman} provide insights into the nature and direction of the relationships between feature parameters and result values. By combining these two methods, we have uncovered notable findings, which will be explored in detail in the following section.

\section{Discussion and conclusions}
\label{sec:conclu}

From the results presented in Figures \ref{im:spearman}, \ref{im:importances1} and \ref{im:importances2} we can derive numerous insights for each result value ($N_{x_1}$, $N_{x_2}$, ... $<A_{sec}>$, and $\Delta A$, see Section \ref{sec:feat_res}). However, we will limit our discussion to what we consider the three most significant findings in the following subsections:

\subsection{Evolution of $x_1$ and $x_2$ orbits and double barred galaxies}
\label{sec:doubleB}

It is essential to note that the methods introduced in Section \ref{sec:feat_res} are directly applied to the response model, allowing for the tracking of changes introduced whenever the parameter combinations depicted in Figure \ref{im:spearman} occur. This approach provides significant practical value in the analysis of fully self-consistent N-body models, enabling a comprehensive assessment of the evolution of key orbital structures, particularly the $x_1$ and $x_2$ families.\\

In Figure \ref{im:importances1}, it is evident that the $\mathcal{R}$ parameter is the second most influential feature for determining $N_{x_1}$. Notably, the correlation coefficient between these two parameters in Figure \ref{im:spearman} is strongly negative. Conversely, the most significant feature for $N_{x_2}$ is $\Omega_B$, which also exhibits a negative correlation.\\

Previous studies using N-body simulations (e.g., \citealp{2003MNRAS.341.1179A}; \citealp{2014MNRAS.438.2201M}) have demonstrated that in barred galaxies, $\Omega_B$ tends to decrease over time, consequently $\mathcal{R}$ increases. When combined with our findings, this suggests that barred galaxies may initially feature a fast bar, characterized by a high number of $x_1$ orbits, and gradually lose some of these orbits as the system evolves. Simultaneously, as $\Omega_B$ decreases, the bar could be acquiring a larger number of $x_2$ orbits.\\

Moreover, the presence of a significant concentration of $x_2$ orbits could offer an explanation for the existence of a secondary bar in certain galaxies (\citealp{1996A&AS..118..461F}; \citealp{2002MNRAS.329..502M}; \citealp{2004A&A...415..941E}; \citealp{2015A&A...575A...7W}; \citealp{2024MNRAS.528.3613E}). This indicates that double-barred galaxies may be dynamically more evolved systems.\\

\subsection{Characterizing bar strength evolution through $\Delta A$}
\label{sec:delA}

Figure \ref{im:A2_OG} presents two distinct bar strength curves over time, $A_2(t)$. The first curve exhibits rapid growth until it reaches a saturation point. After saturation, the bar rapidly weakens until it reaches an equilibrium (as expected for response models) with $<A_{sec}>$. In contrast, in the second curve the bar also weakens beyond the saturation point, although at a much slower rate. Looking in the other $A(t)$ curves we can note there are intermediate cases.\\

The parameter $\Delta A$ plays a crucial role in understanding this behavior. A high positive value of $\Delta A$ indicates that $A_{sat}$ is larger than $<A_{sec}>$, aligning with the behavior observed in the first case in Figure \ref{im:A2_OG}. Conversely, a low $\Delta A$ corresponds to a behavior more related to the second case shown in the same figure.\\

As previously discussed, the $x_1$ family of orbits forms the backbone of rotating Ferrers bars, remaining stable in the region that supports the bar. In a response model, the decreasing of $\Delta A$ could be attributed to the presence of chaotic or escape orbits within the system. To verify this last statement, a study using an index to quantify the chaotic behavior of the orbits as GALI2 (\citealp{2007PhyD..231...30S}; \citealp{2017ApJ...850..145C}; \citealp{2019RMxAA..55..321C}) could be performed. However, this is beyond the scope of the present paper.\\

The permutation feature importance analysis for $\Delta A$ (Figure \ref{im:importances2}) shows that the primary feature parameter, with a significant larger importance compared to other feature parameters, is $\mathcal{R}$. Furthermore, the Spearman correlation between $\Delta A$ and $\mathcal{R}$ (as shown in Figure \ref{im:spearman}) is negative. From this, we can infer the following:\\

\begin{itemize}
    \item \textbf{Slow Bars:} In cases of slow bars, the models loose $x_1$ particles at a very slow rate after reaching the saturation point (as in the second case mentioned earlier in Figure \ref{im:A2_OG}).

    \item \textbf{Fast Bars:} Conversely, fast bars experience high particle loss beyond the saturation point (similar to the first case mentioned previously in Figure \ref{im:A2_OG}).\\
\end{itemize}

\subsection{Impact of Disk-to-Halo ratio on bar formation}
\label{sec:DHratio}

In subsection \ref{sec:GM}, we constructed three distinct axisymmetric models. These models exhibit the closest circular velocity to that of the Milky Way within the first 40 kpc, as shown in Figure \ref{im:vel_circ}. Despite exhibiting degeneracy in terms of circular velocity, discernible differences exist among them, primarily in the disk-to-halo ratio. Model 1 shows a strong disk influence compared to the halo in $R<8$ kpc, whereas Model 3 shows a stronger halo influence compared to the disk from $R>3$ kpc. Model 2 represents an intermediate case.\\

Since earlier works as \citet{1986MNRAS.221..213A}, it is known that galaxies with a stronger disk influence, as seen in Model 1, tend to form bars more rapidly than those in which the halo is predominant, like Model 3 (see also \citet{2023MNRAS.525.3162V}). However, the permutation importance for all eight result values shown in Figures \ref{im:importances1} and \ref{im:importances2} indicate that the variations among the axisymmetric models have a negligible impact on the result values. This finding is corroborated by the Spearman correlation coefficients presented in Figure \ref{im:spearman}, where it is evident that most of the coefficients related to the axisymmetric model are nearly zero. The most significant correlation is for $\Delta A$ with a coefficient of $-0.20$, which is insufficient to establish a strong anticorrelation.\\

The apparent contradiction presented here arises from the context of this work, which belongs to rigid potential models. It is noteworthy that when one performs a N-body fully self-consistent models, the structural parameters of the galactic components exhibit temporal evolution, and there is a transfer of angular momentum among the components. In our research, we are imposing the same bar model characterized by parameters $\mu_B$, $a_B$, and $\Omega_B$ across all three axisymmetric models.\\

Consequently, we can conclude that for degenerate models, using rigid potentials, the variation in disk-to-halo does not significantly affect the the formation of $x_1$ and $x_2$ orbits.\\

Finally, it is crucial to highlight that combining the Spearman correlation coefficients with the feature importance derived from a Random Forest Regressor significantly enhances the analysis of the effects of different input parameters on output results. The Spearman correlation provides insight into the monotonic relationships between values, while the feature importance in a Random Forest Regressor evaluates the overall significance of each parameter. Using both methods allows for a comprehensive understanding of the importance and behavior of various parameters.\\

In conclusion, our work provides a detailed investigation into the dynamics of barred galaxies, offering insights into the interplay between various galactic parameters and the formation and evolution of galactic structures. The methodologies employed and the findings derived from this study contribute to the broader understanding of galactic dynamics and serve as a foundation for future research in this field.

\section{Acknowledgement}
\label{sec:Acknowl}
We sincerely thank the referee for his/her comments which have greatly improved and clarified the presentation of our study.

\end{document}